\documentclass[aps,pra,twocolumn,showpacs,superscriptaddress]{revtex4} 
\usepackage{graphicx}
\usepackage{amsmath}
\usepackage{amssymb}
\usepackage{epstopdf}
\usepackage{amsfonts}
\usepackage{dcolumn}
\begin{document}
\title{Energy spectrum of a harmonically trapped two-atom system with spin-orbit
coupling}
\author{Q. Guan, X. Y. Yin, Seyed Ebrahim Gharashi, and D. Blume}
\address{Department of Physics and Astronomy,
Washington State University,
  Pullman, Washington 99164-2814, USA}
\date{\today}

\begin{abstract}
Ultracold atomic gases provide
a novel platform with which to study spin-orbit
coupling, a mechanism that plays a central role in the
nuclear shell model, atomic fine structure and two-dimensional
electron gases.
This paper introduces a theoretical framework that allows for the
efficient determination of the eigenenergies and eigenstates
of a harmonically trapped two-atom system with short-range interaction
subject to an equal mixture of Rashba and Dresselhaus spin-orbit coupling
created through Raman coupling of atomic hyperfine states.
Energy spectra for experimentally relevant parameter combinations are
presented and future extensions of the approach are discussed.
\end{abstract}
\maketitle

Over the past decade, much progress has been made in preparing
isolated ultracold few-atom systems 
experimentally~\cite{bloch10,Jochim1,Jochim3,regal}.
Moreover, a variety of tools
for manipulating and probing such systems have been developed.
On the theoretical side, a number of analytical
and numerical approaches have been 
developed~\cite{braatenReview,BlumeReview,ritt11,Busch98,Idziaszek05,Idziaszek06,kestner07,blume07,vankolck07}.
A large number of analytical treatments approximate the 
true alkali atom-alkali atom potential by a zero-range
potential~\cite{Busch98,ferm34,Huang57,HuangText}. 
This replacement yields reliable results
in the low-energy regime where the de Broglie wave length is larger 
than the van der Waals length.
For example,
using zero-range contact interactions, the energy spectrum of 
two harmonically trapped atoms has been determined 
analytically~\cite{Busch98,Idziaszek05,Idziaszek06}.
These two-body solutions are available in 1D, 2D and 3D~\cite{Busch98}, 
and have 
played a vital role in guiding and interpreting 
experiments~\cite{esslinger05,esslinger06,jochim12}
as well as in theoretical studies of the two-body 
dynamics~\cite{blume03,bohn13}
and of larger harmonically trapped 
systems~\cite{kestner07,blume07,vankolck07,HuiPRA,Gharashi12,Gharashi13}.

Recently, synthetic gauge fields, which allow for the realization of 
Hamiltonians that contain spin-orbit coupling terms, have been
realized 
experimentally~\cite{rmp11,spielmannat13,spielmanreview13,zhaireview,spielman11,spielman13,jzhang12,jzhang13,zwierlein12,engels13,greene13}.
The purpose of this paper is to address how the trapped two-particle
spectrum, obtained by modeling the two-body interaction
via a zero-range $\delta$-function, 
changes in the presence of spin-orbit and Raman coupling.
While the two-particle system with spin-orbit coupling
in free space~\cite{magarill06,shenoy11,galitski12,sthbypu}
as well as the 
trapped single-particle system with spin-orbit 
coupling~\cite{zinner12,anderson2013}  
have received
considerable attention, little is known
about the trapped two-particle system 
with spin-orbit coupling and two-body interaction~\cite{desmondSOC,puPRA13}.
In going from the trapped single-atom to the
trapped two-atom system, a new length scale, i.e., 
the atom-atom scattering length, comes into play.
Thus, an interesting question concerns the interplay
between the interaction energy and the energy scales associated 
with the spin-orbit and Raman coupling strengths.

Our framework applies to the situation where the spin-orbit 
(or more precisely, spin-momentum)
coupling term 
acts, as in recent 
experiments~\cite{spielman11,spielman13,jzhang12,jzhang13,zwierlein12,engels13,greene13}, 
along one spatial direction,
say the $x$-direction. 
This corresponds to an equal mixture of Rashba 
and Dresselhaus spin-orbit coupling~\cite{Rashba,Dresselhaus}.
For 
simplicity, we assume that the harmonic confinement in the other two
spatial directions is much larger than that
in the direction where the
spin-orbit coupling term acts. This assumption reduces the problem to an
effective one-dimensional Hamiltonian in the $x$-coordinates with 
effective 1D two-body interaction.
The relationship between the true 3D atom-atom and effective 1D atom-atom
interaction has been derived in Refs.~\cite{Olsh98,olshanii2,chinesepaper}.
We find analytical solutions to the two-atom system for arbitrary 
spin-orbit coupling strength and scattering length and vanishing
Raman coupling strength. 
The case of non-zero Raman coupling strength is treated by expanding the
system Hamiltonian in terms of the eigenstates for vanishing 
Raman coupling strength. We find that the relevant 
Hamiltonian matrix elements have closed analytical expressions, leaving the
matrix diagonalization as the only numerical step.
The developed framework can, as discussed toward the end of our paper, 
be readily generalized to a spherically-symmetric harmonic trap
or an axi\-symmetric trap.
Moreover, the framework developed also lays the
groundwork for treating dynamical
aspects of trapped two-body systems with non-vanishing spin-orbit
and Raman
coupling strengths and for treating the corresponding 
three-body system.

We consider two structureless one-dimensional particles of mass $m$
subject to a single-particle spin-orbit coupling term of strength
$k_{\text{so}}$, a Raman coupling term with strength
$\Omega$, detuning $\delta$, and an external harmonic potential with
angular trapping frequency $\omega$.
For $k_{\text{so}}=\Omega=\delta=0$,
the two-particle Hamiltonian is given by 
$H_{\text{sr}}$,
\begin{eqnarray}
H_{\text{sr}} = 
\sum_{j=1}^2
\left(
\frac{-\hbar^2}{2m} \frac{\partial^2}{\partial x_j^2}
+ \frac{1}{2} m \omega^2 x_j^2 \right)
+ V_{2\text{b}} (x_1-x_2),
\end{eqnarray}
where $x_j$ denotes the position coordinate of
the $j$th particle and $V_{2\text{b}}$ the short-range interaction potential.
For non-zero $k_{\text{so}}$, $\Omega$ and $\delta$,
the two-particle
Hamiltonian is given by $H$,
\begin{eqnarray}
H = H_{\text{sr}} \hat{I} +
\sum_{j=1}^2 
\left[
\frac{\hbar k_{\text{so}}}{m}  p_{xj} \sigma_y^{(j)} +
\frac{\Omega}{2} \sigma_{x}^{(j)}
+\frac{\delta}{2} \sigma_{y}^{(j)}
 \right],
\end{eqnarray}
where $\sigma_x^{(j)}$  and $\sigma_y^{(j)}$ 
denote
Pauli matrices, $\hat{I}$ the identity matrix and
$p_{xj}$ the momentum of the $j$th particle.
In the following, we first derive solutions to the 
time-independent Schr\"odinger
equation governed by $H$ with $\Omega=0$ and then discuss
how to obtain the solutions for non-zero $\Omega$.

To determine the eigenstates and eigenenergies of $H$, we perform
a rotation in spin space~\cite{sakurai}.
Specifically, we define $\tilde{H}$ via a unitary transformation
of $H$,
$\tilde{H}= U^{\dagger} H U$, where
$U = \exp[\imath (\sigma_x^{(1)}+\sigma_x^{(2)}) \pi/4]$.
The eigenenergies of the Hamiltonian $H$ and 
$\tilde{H}$ coincide while the eigenstates
$\Psi$ of
the Hamiltonian $H$ are related to the eigenstates $\tilde{\Psi}$ of
the Hamiltonian 
$\tilde{H}$ through
$\Psi = U \tilde{\Psi}$.
A straightforward calculation shows that
$U^{\dagger} \sigma_q^{(j)} U = \sigma_x^{(j)}$ and
$\sigma_z^{(j)}$ for 
$q=x$ and $y$, respectively.
Correspondingly, we have
\begin{eqnarray}
\tilde{H} = 
H_{\text{sr}} \hat{I} + \sum_{j=1}^2 \left[
\frac{\hbar k_{\text{so}}}{m} p_{xj} \sigma_z^{(j)}
+ \frac{\Omega}{2} \sigma_x^{(j)} + \frac{\delta}{2} \sigma_z^{(j)} 
\right].
\end{eqnarray}

\begin{figure}
\centering
\includegraphics[angle=0,width=0.4\textwidth]{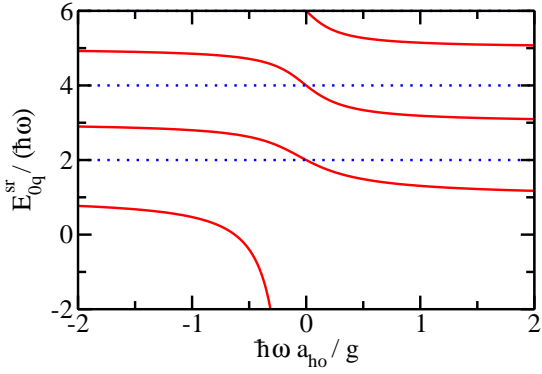}
\vspace*{0.5in}
\caption{(Color online)
Zero-range energies $E_{0q}^{\text{sr}}$
as a function of $\hbar \omega a_{\text{ho}} /g$.
Solid and dotted lines show the energies corresponding to
states that are even and odd, respectively, in the relative coordinate.
 }\label{fig1}
\end{figure}

For $\Omega=0$, $\tilde{H}$ is diagonal in the pseudo-spin basis
$| \uparrow \rangle_1 | \uparrow \rangle_2$,
$| \uparrow \rangle_1 | \downarrow \rangle_2$,
$| \downarrow \rangle_1 | \uparrow \rangle_2$ and
$| \downarrow \rangle_1 | \downarrow \rangle_2$
with diagonal elements
$\tilde{H}^{\uparrow \uparrow}$,
$\tilde{H}^{\uparrow \downarrow}$,
$\tilde{H}^{\downarrow \uparrow}$ and
$\tilde{H}^{\downarrow \downarrow}$.
To find the corresponding eigenstates, we approximate the
two-body interaction by a delta-function interaction
with coupling constant $g$,
$V_{2\text{b}}(x_1-x_2)=g \delta(x_1-x_2)$.
For this interaction model,
the eigenenergies and eigenstates of $H_{\text{sr}}$ 
are known in compact form~\cite{Busch98}.
States that are even in the relative coordinate are affected by
the coupling constant $g$ while those that are odd in the relative coordinate
are not.
For states that are even in the relative coordinate, 
the eigenenergies $E_{nq}^{\text{sr}}$ of $H_{\text{sr}}$ 
(see solid lines in Fig.~\ref{fig1} for the $n=0$ energies) are given by
$(n+2q+1) \hbar \omega$, 
where the center of mass
quantum number $n$ takes the values $n=0,1,\cdots$ and the non-integer
quantum number $q$ is determined by the transcendental equation~\cite{Busch98}
\begin{eqnarray} 
\frac{2\Gamma(-q+1/2)}{\Gamma(-q)} = 
-\frac{g}{\sqrt{2} \hbar \omega a_{\text{ho}}};
\end{eqnarray}
here, $a_{\text{ho}}$ denotes the harmonic
oscillator length, $a_{\text{ho}} = \sqrt{\hbar/(m \omega)}$.
The corresponding eigenfunctions 
$\psi_{nq}^{\text{sr}}(x,X)$ are 
given by 
$\phi_q(x) \Phi_n(X)$, where 
the relative and center of mass coordinates are defined through
$x=(x_1-x_2)/\sqrt{2}$ and $X=(x_1+x_2)/\sqrt{2}$,
respectively.
The relative wave functions $\phi_q(x)$ can be written
in terms of the confluent hypergeometric function $U$~\cite{Busch98},
$\phi_q(x)= N_q U(-q,1/2,(x/a_{\text{ho}})^2) \exp[-x^2/(2 a_{\text{ho}}^2)]$,
where $N_q$ denotes 
a
normalization constant.
The center of mass functions $\Phi_n(X)$ are given by the
one-dimensional harmonic oscillator functions for a mass $m$ particle,
$\Phi_n(X)=N_n^{\text{ni}} H_n(X/a_{\text{ho}}) 
\exp[-X^2/(2 a_{\text{ho}}^2)]$,
where $H_n$ denotes the Hermite polynomial of order $n$
and $N_n^{\text{ni}}=(\sqrt{\pi} 2^n n! a_{\text{ho}})^{-1/2}$.
For states that are odd in the relative coordinate, 
the eigenenergies $E_{nq}^{\text{sr}}$ of $H_{\text{sr}}$ 
(see dotted lines in Fig.~\ref{fig1} for the $n=0$ energies) are given by
$(n+2q+2) \hbar \omega$, where $q$ and $n$ take the values $0,1,\cdots$.
In this case, the eigenfunctions $\psi_{nq}^{\text{sr}}(x,X)$ are
simply products of 
the non-interacting harmonic oscillator functions in $x$ and $X$.

In addition to using the known properties of $H_{\text{sr}}$, 
we take advantage of the
fact
that the kinetic energy
$(p_{x1}^2 + p_{x2}^2)/(2m)$ of $H_{\text{sr}}$
and the
$k_{\text{so}}$ dependent terms
can be combined,
\begin{eqnarray}
\frac{p_{xj}^2}{2m} \pm \frac{\hbar k_{\text{so}}}{m} p_{xj} =
\frac{(p_{xj} \pm \hbar k_{\text{so}})^2}{2m} - 
\frac{\hbar^2 k_{\text{so}}^2}{2m}.
\end{eqnarray}
This identity suggests that the momentum-dependent spin-orbit
coupling terms add a ``momentum boost'' 
to the solutions 
$\psi_{nq}^{\text{sr}}(x,X)$
of $H_{\text{sr}}$.
Indeed, it is readily verified  that the eigenstates
of 
$\tilde{H}^{\uparrow \uparrow}$,
$\tilde{H}^{\uparrow \downarrow}$,
$\tilde{H}^{\downarrow \uparrow}$, and
$\tilde{H}^{\downarrow \downarrow}$ are
given by
\begin{eqnarray}
\label{eq_psi1}
\tilde{\psi}_{nq}^{\uparrow \uparrow}(x,X)= 
\exp(- \imath \sqrt{2}  k_{\text{so}}X)
\psi_{n q}^{\text{sr}}(x,X),
\end{eqnarray}
\begin{eqnarray}
\label{eq_psi2}
\tilde{\psi}_{nq}^{\uparrow \downarrow}(x,X)= 
\exp(- \imath \sqrt{2}  k_{\text{so}}x)
\psi_{n q}^{\text{sr}}(x,X), 
\end{eqnarray}
\begin{eqnarray}
\label{eq_psi3}
\tilde{\psi}_{nq}^{\downarrow \uparrow}(x,X)= 
\exp( \imath \sqrt{2}  k_{\text{so}}x)
\psi_{n q}^{\text{sr}}(x,X),
\end{eqnarray}
and
\begin{eqnarray}
\label{eq_psi4}
\tilde{\psi}_{nq}^{\downarrow \downarrow}(x,X)= 
\exp( \imath \sqrt{2}  k_{\text{so}}X)
\psi_{n q}^{\text{sr}}(x,X),
\end{eqnarray}
respectively.
For fixed $g$ and $n$ and vanishing $\delta$, the states 
given in Eqs.~(\ref{eq_psi1})-(\ref{eq_psi4}) are degenerate
with eigenenergies $E_{nq}=E_{nq}^{\text{sr}} - \hbar^2 k_{\text{so}}^2/m$.
For $|g| = \infty$, the degeneracy doubles 
(see the crossings of the solid and dotted lines in Fig.~\ref{fig1})
since $q$
takes the values
$1/2,3/2,\cdots$ for $\psi_{nq}^{\text{sr}}$ 
that are even in $x$ and the values 
$0,1,2,\cdots$ for $\psi_{nq}^{\text{sr}}$  that are odd in 
$x$, i.e., since each of the $\psi_{nq}^{\text{sr}}$ odd in $x$
is degenerate with one of the $\psi_{nq}^{\text{sr}}$ even in $x$.
For non-vanishing $\delta$, the energies are shifted by
$\delta$, $0$, $0$ and $-\delta$, respectively.
The eigenenergies are simply the sum of a term
that depends on the coupling constant $g$, a center of mass contribution
that is characterized by $n$, a term that depends on the 
square of the spin-orbit coupling
strength $k_{\text{so}}$ and a term that depends on the detuning $\delta$.

If $\Omega$ is non-zero, the Hamiltonian $\tilde{H}$ expressed in
the 
$| \uparrow \rangle_1 | \uparrow \rangle_2$, 
$| \uparrow \rangle_1 | \downarrow \rangle_2$, 
$| \downarrow \rangle_1 | \uparrow \rangle_2$, 
and 
$| \downarrow \rangle_1 | \downarrow \rangle_2$ 
pseudo-spin basis is no longer
diagonal.
To determine the eigenenergies and eigenstates for non-zero $\Omega$, we
expand $\tilde{H}$ in terms of the eigenstates
$\tilde{\psi}_{n q}^{\sigma_1 \sigma_2}$, where $\sigma_1$ and $\sigma_2$ 
take the values $\uparrow$ and $\downarrow$.
The
off-diagonal matrix elements 
$H_{n'q',nq}^{\sigma_1' \sigma_2',\sigma_1 \sigma_2}$,
\begin{eqnarray}
\label{eq_offdiagonal}
 H_{n'q',nq}^{\sigma_1' \sigma_2',\sigma_1 \sigma_2}
=
\frac{\Omega}{2} 
\int_{-\infty}^{\infty} \int_{-\infty}^{\infty}
(\tilde{\psi}_{n'q'}^{\sigma_1' \sigma_2'})^*
\tilde{\psi}_{nq}^{\sigma_1 \sigma_2}
dx dX,
\end{eqnarray}
can be separated into two
one-dimensional integrals,
\begin{eqnarray}
 H_{n'q',nq}^{\sigma_1' \sigma_2',\sigma_1 \sigma_2}
=
\frac{ \Omega}{2} 
I_{q'q}^{\sigma_1' \sigma_2',\sigma_1 \sigma_2} J_{n'n}^{\sigma_1' \sigma_2',\sigma_1 \sigma_2}, 
\end{eqnarray}
where
\begin{eqnarray}
I_{q'q}^{\sigma_1' \sigma_2',\sigma_1 \sigma_2} = \int_{-\infty}^{\infty} 
\exp( \eta \imath \sqrt{2} k_{\text{so}} x) [\phi_{q'}(x)]^* \phi_q(x) dx
\end{eqnarray}
and
\begin{eqnarray}
J_{n'n}^{\sigma_1' \sigma_2',\sigma_1 \sigma_2} 
= \int_{-\infty}^{\infty} 
\exp( \xi \imath \sqrt{2} k_{\text{so}} X) [\Phi_{n'}(X)]^* \Phi_n(X) dX.
\end{eqnarray}
The sign of the exponent is determined by the pseudo-spin combinations:
$(\eta,\xi)=(-,+)$, $(+,+)$, $(-,-)$ and $(+,-)$
for $(\sigma_1' \sigma_2',\sigma_1 \sigma_2)=
(\uparrow \uparrow,\uparrow \downarrow)$,
$(\uparrow \uparrow,\downarrow \uparrow)$,
$(\downarrow \downarrow,\uparrow \downarrow)$ and
$(\downarrow \downarrow,\downarrow \uparrow)$,
respectively.

The integral 
$J_{n'n}^{\sigma_1' \sigma_2',\sigma_1 \sigma_2} $
over the center of mass coordinate 
coincides with the Fourier transform
of the product of two one-dimensional harmonic oscillator
eigenstates.
For $\xi=\pm$ ($n' \le n$),
we find~\cite{integraltable}
\begin{eqnarray}
\label{eq_cmfinal}
J_{n'n}^{\sigma_1' \sigma_2',\sigma_1 \sigma_2} =
a_{\text{ho}}
\sqrt{\pi} N_{n'}^{\text{ni}} N_n^{\text{ni}}n'!
2^{(n+n')/2} 
\nonumber \\
\times
L_{n'}^{(n-n')}((a_{\text{ho}} k_{\text{so}})^2)
(\pm \imath k_{\text{so}} a_{\text{ho}})^{n-n'}
\exp[-(k_{\text{so}} a_{\text{ho}})^2/2] ,
\end{eqnarray}
where $L_n^{(n-n')}$ denotes the associated Laguerre polynomial.

The integral 
$I_{q'q}^{\sigma_1' \sigma_2',\sigma_1 \sigma_2}$
over the relative coordinate can be performed by expanding 
$[\phi_{q'}(x)]^*$ and $\phi_{q}(x)$ in terms of the non-interacting
harmonic oscillator functions $\phi_l^{\text{ni}}(x)$,
$\phi_{q}(x)=\lim_{l_{\text{max}}\rightarrow \infty}
\sum_{l=0}^{l_{\text{max}}} c_l^{(q)} \phi_l^{\text{ni}}(x)$,
where the expansion coefficients $c_l^{(q)}$ can be obtained
analytically~\cite{Busch98}.
The integral $I_{q'q}^{\sigma_1' \sigma_2', \sigma_1 \sigma_2}$
then becomes a double sum over integrals that have the same structure as
the center of mass integrals
$J_{n'n}^{\sigma_1' \sigma_2', \sigma_1 \sigma_2}$.
In the calculations reported below, we use
a finite cutoff $l_{\text{max}}$. The ``optimal'' cutoff depends on the value
of $g$ considered, the number of relative functions $\phi_q(x)$ included
in the basis and the desired accuracy.
For $|g|=\infty$, we find,
as in the $g=0$ case, a closed analytical expression for
the integral 
$I_{q'q}^{\sigma_1' \sigma_2', \sigma_1 \sigma_2}$.
Having analytical expressions for the matrix elements of 
$\tilde{H}$, the eigenenergies can be obtained through matrix
diagonalization.

To obtain basis functions with good
quantum numbers, we work with linear
combinations of the
functions given in Eqs.~(\ref{eq_psi1})-(\ref{eq_psi4}),
i.e., we work with the basis functions 
$\tilde{\psi}_{X,\pm}=
(\tilde{\psi}_{nq}^{\uparrow \uparrow} | \uparrow \rangle_1 |\uparrow \rangle_2
\pm \tilde{\psi}_{nq}^{\downarrow \downarrow} | \downarrow \rangle_1 | \downarrow \rangle_2) /\sqrt{2}$
and
$\tilde{\psi}_{x,\pm}=
(\tilde{\psi}_{nq}^{\uparrow \downarrow} | \uparrow \rangle_1 | \downarrow \rangle_2
\pm \tilde{\psi}_{nq}^{\downarrow \uparrow} | \downarrow \rangle_1 | \uparrow \rangle_2) /\sqrt{2}$.
By properly combining the parts of $\psi_{nq}^{\text{sr}}$
that are even or odd in the relative coordinate and even or odd in
the center of mass coordinate, we construct basis
functions that are eigenstates of the operators
$P_{12}$ and $Y_{12}$.
The operator $P_{12}$ exchanges the coordinates (position and spin)
of particles 1 and 2. Basis functions that are unchanged under
the operation $P_{12}$ are needed to describe states with bosonic symmetry
$(p_{12}=+1$)
and those that pick up
a minus sign under the operation $P_{12}$ are needed to describe states with
fermionic symmetry ($p_{12}=-1$).
The operator $Y_{12}$ can be written as 
$\sigma_{x}^{(1)}\sigma_{x}^{(2)} P P_{12}$, where
the parity operator $P$ changes $x_j$ to $-x_j$ ($j=1$ and $2$).
The $Y_{12}$ operator determines the ``helicity'' of the system.
We label the eigenstates by $(p_{12},y_{12})$, where 
$p_{12}=\pm 1$ and
$y_{12}=\pm1$ are defined by their actions on an eigenstate.
The basis functions with $(+1,+1)$ symmetry, for example, are
given 
by $\tilde{\psi}_{X,+}$ with $\phi_q(x)$ even and $\Phi_n(X)$ even,
by $\tilde{\psi}_{X,-}$ with $\phi_q(x)$ even and $\Phi_n(X)$ odd,
by $\tilde{\psi}_{x,+}$ with $\phi_q(x)$ even and $\Phi_n(X)$ even,
and
by $\tilde{\psi}_{x,-}$ with $\phi_q(x)$ odd and $\Phi_n(X)$ even.
\begin{widetext}

\begin{figure}
\centering
\includegraphics[angle=0,width=0.8\textwidth]{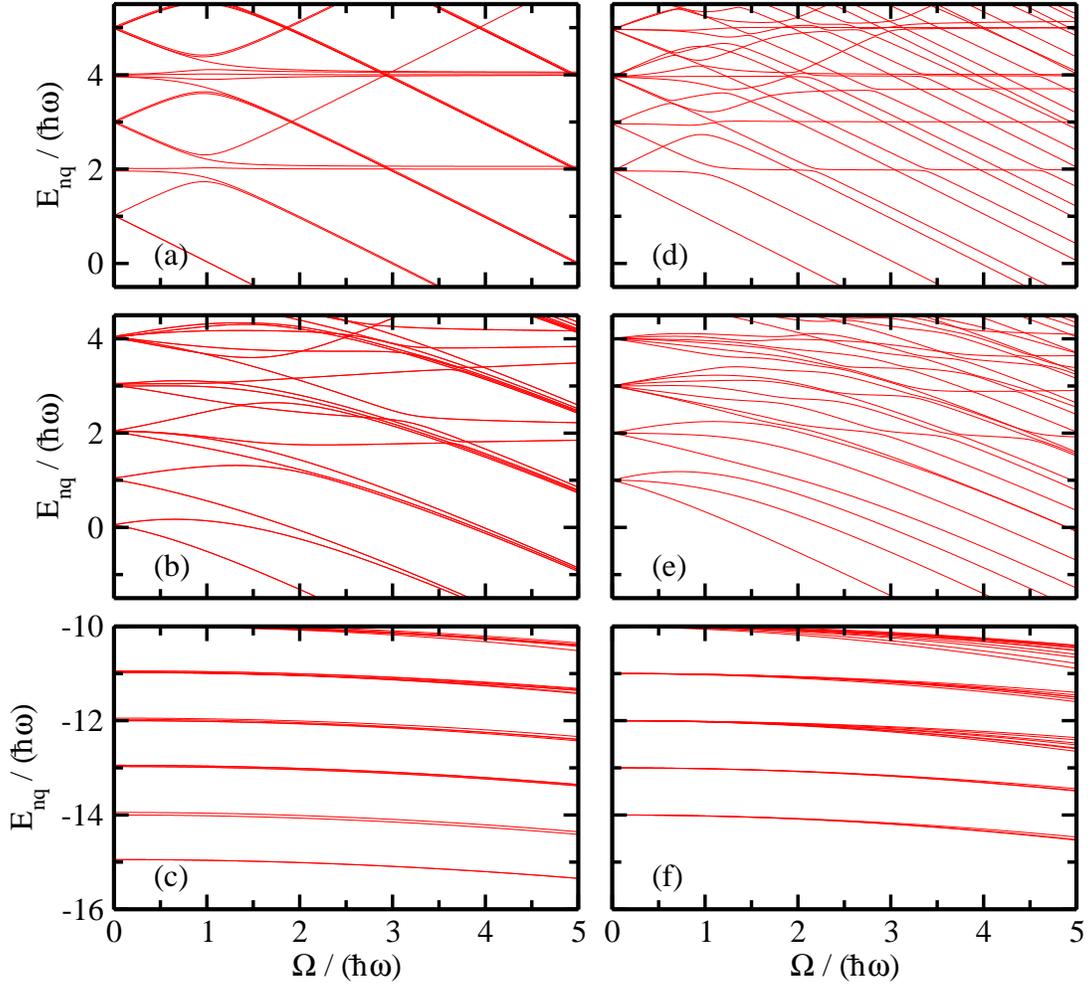}
\vspace*{0.1in}
\caption{(Color online)
Eigenenergies $E_{nq}$ corresponding to eigenstates
with $(p_{12},y_{12})=(+1,+1)$ 
as a function of 
$ \Omega$ for 
$\delta=0$,
and 
(a) $a_{\text{ho}} k_{\text{so}} = 0.2$ and 
$g=a_{\text{ho}} \hbar \omega/\sqrt{50}$;
(b) $a_{\text{ho}} k_{\text{so}} = 1$ and
$g=a_{\text{ho}} \hbar \omega/\sqrt{50}$;
(c) $a_{\text{ho}} k_{\text{so}} = 4$ and
$g=a_{\text{ho}} \hbar \omega/\sqrt{50}$;
(d) $a_{\text{ho}} k_{\text{so}} = 0.2$ and 
$|g|=\infty$;
(e) $a_{\text{ho}} k_{\text{so}} = 1$ and
$|g|=\infty$;
and
(f) $a_{\text{ho}} k_{\text{so}} = 4$ and
$|g|=\infty$,
respectively.
 }\label{fig_ensoc}
\end{figure} 

\end{widetext}

As an example, Figs.~\ref{fig_ensoc}(a)-\ref{fig_ensoc}(c)
show energy spectra corresponding to eigenstates
with $(p_{12},y_{12})=(+1,+1)$ as a function of the
Raman coupling strength $\Omega$ for vanishing detuning $\delta$, 
small coupling constant $g$, 
$g\approx 0.1414 a_{\text{ho}} \hbar \omega$,
and three different spin-orbit coupling strengths $k_{\text{so}}$.
The energy spectrum in Fig.~\ref{fig_ensoc}(a) is, to leading order,
given 
by the spectrum for $\delta=k_{\text{so}}=g=0$.
In this limiting case, the energies
are equal to $(2 j +1)\hbar \omega \pm  \Omega$
and $2j \hbar \omega$, where $j=0,1,\cdots$.
Finite $g$ and $k_{\text{so}}$ values
introduce shifts and avoided crossings.
Specifically,
the small positive coupling constant $g$ 
introduces a positive energy shift for the states
that are even in the relative coordinate, which---in first-order perturbation
theory---is given by 
$\frac{1}{\sqrt{2\pi}}\frac{(2q)!}{(q!)^24^q}g/a_{\text{ho}}$. 
The spin-orbit coupling term introduces, in the small $\Omega$ regime,
a small down shift that is proportional to $k_{\text{so}}^2$.
This down shift is negligible in Fig.~\ref{fig_ensoc}(a)
but clearly visible in Figs.~\ref{fig_ensoc}(b) and \ref{fig_ensoc}(c).
Moreover, the spin-orbit coupling introduces avoided crossings.
The broadest avoided crossings occur around
$\Omega=\hbar \omega$, where states with the same $q$ but
$n$ quantum numbers  
that differ by one are coupled. 
The reason is that the spin-orbit coupling
term connects, via the total momentum operator, 
states in first-order perturbation theory 
if the states' $n$ quantum numbers differ by one
and in higher-order perturbation theory otherwise.

Figures~\ref{fig_ensoc}(d)-\ref{fig_ensoc}(f) 
show energy spectra for the strong coupling
limit, i.e., for $|g| \rightarrow \infty$, as a function
of the Raman coupling $\Omega$ for vanishing detuning $\delta$.
To facilitate the comparison between the
large and small $g$ limits,
the spin-orbit coupling strengths $k_{\text{so}}$
in Figs.~\ref{fig_ensoc}(d)-\ref{fig_ensoc}(f) 
are the same as
in Figs.~\ref{fig_ensoc}(a)-\ref{fig_ensoc}(c).
In the regime where $\Omega \ll \hbar^2 k_{\text{so}}^2/m$,
the energies change approximately linearly with $\Omega$
(with positive, vanishing or negative slope). 
When $ \Omega \gg \hbar^2 k_{\text{so}}^2/m$, the low-lying portion
of the energy spectrum consists of approximately parallel energy levels that
can be parameterized as  $c - \Omega$, where $c$ is a constant.

As already aluded to in the introduction,
the theoretical framework developed can be generalized to higher-dimensional
trapping geometries.
For a spherically symmetric 3D system, e.g., the eigenstates
of the 3D Hamiltonian $H_{\text{sr}}$
with 3D contact interaction can be expanded
in terms of products of 2D and 1D harmonic oscillator states
using cylindrical coordinates.
As in the 1D case pursued in this work,
the matrix elements for the higher-dimensional
system can be calculated analytically.
Axisymmetric harmonic traps with spin-orbit coupling in one
direction
can be treated analogously.
Furthermore, using the eigenstates of the trapped three-particle system 
in 1D,
2D or 3D with contact 
interactions~\cite{kestner07,HuiPRA,Gharashi12} and expressing the
three-particle Hamiltonian in terms of the eight
pseudo-spin states,
a non-zero $\Omega$ introduces 
off-diagonal elements that can be calculated analytically
following steps similar to those discussed in this paper.

Summarizing,
this work introduced a theoretical framework
that allows for the efficient determination of the
energy spectrum and eigenstates of the trapped two-particle system
in 1D with contact interaction
and spin-orbit and Raman coupling terms.
The energy spectra show a rich dependence
on the interaction, spin-orbit and Raman coupling strengths.
The framework presented provides an important stepping stone
for treating more complicated systems with spin-orbit coupling,
such as higher-dimensional
two-body systems and three-body systems.

Acknowledgement:
Support by the National Science Foundation  through grant number
PHY-1205443
and insightful discussions with P. Engels are
 gratefully acknowledged.
DB and XYY acknowledge support from the Institute for Nuclear Theory
during the program  INT-14-1,
``Universality in Few-Body Systems: Theoretical
Challenges and New Directions''.



\begin{thebibliography}{10}

\bibitem{bloch10}
S. Will, T. Best, U. Schneider, L. Hackerm\"uller, D.-S. L\"uhmann, and
I. Bloch,
Nature {\bf{465}}, 197 (2010).


\bibitem{Jochim1}
F.~Serwane, G.~Z\"urn, T.~Lompe, T.~B.~Ottenstein, A.~N.~Wenz, and S.~Jochim,
Science {\bf{332}}, 6027 (2011).


\bibitem{Jochim3}
A.~N.~Wenz, G.~Z\"urn, S.~Murmann, I.~Brouzos, T.~Lompe, and S.~Jochim,
Science {\bf{342}}, 457 (2013).

\bibitem{regal}
A. M. Kaufman, B. J. Lester, C. M. Reynolds, M. L. Wall,
M. Foss-Feig, K. R. A. Hazzard, A. M. Rey, and C. A. Regal,
arXiv:1312.7182.

\bibitem{braatenReview}
E. Braaten and H.-W. Hammer.
Phys. Rep. {\bf{428}}, 259 (2006).

\bibitem{BlumeReview}
D.~Blume,
Rep.~Prog.~Phys. {\bf{75}}, 046401 (2012).

\bibitem{ritt11}
S. T. Rittenhouse, J. von Stecher, J. P. D'Incao, 
N. P. Mehta, and C. H. Greene.
J. Phys. B {\bf{44}}, 172001 (2011).


\bibitem{Busch98}
T. Busch, B.-G. Englert, K. Rz\c a\.zewski, and M. Wilkens,
Found. Phys. {\bf{28}}, 549 (1998).


\bibitem{Idziaszek05}
Z.~Idziaszek and T.~Calarco,
Phys. Rev. A {\bf{71}}, 050701(R) (2005).

\bibitem{Idziaszek06}
Z.~Idziaszek and T.~Calarco,
Phys. Rev. A {\bf{74}}, 022712 (2006).

\bibitem{kestner07}
J. P. Kestner and L.-M. Duan,
Phys. Rev. A {\bf{76}}, 033611 (2007).


\bibitem{blume07}
J. von Stecher, C. H. Greene, and D. Blume,
Phys. Rev. A {\bf{76}}, 053613 (2007).

\bibitem{vankolck07}
I. Stetcu, B. R. Barrett, U. van Kolck, and J. P. Vary,
Phys. Rev. A {\bf{76}}, 063613 (2007).


\bibitem{ferm34}
E. Fermi, Nuovo Cimento {\bf 11},  157  (1934).

\bibitem{Huang57}
K.~Huang and C.~N.~Yang,
Phys. Rev. {\bf{105}}, 767 (1957).

\bibitem{HuangText}
K. Huang, \emph{Statistical Mechanics, 2nd Ed.}
(John Wiley and Sons, Inc., New York, 1963).



\bibitem{esslinger05}
H. Moritz, T. St\"oferle, K. G\"unter, M. K\"ohl,
and T. Esslinger,
Phys. Rev. Lett. {\bf{94}}, 210401 (2005).

\bibitem{esslinger06}
T. St\"oferle, H. Moritz, K. G\"unter, M. K\"ohl,
and T. Esslinger,
Phys. Rev. Lett. {\bf{96}}, 030401 (2006).

\bibitem{jochim12}
G. Z\"urn, F. Serwane, T. Lompe, A. N. Wenz, M. G. Ries, J. E. Bohn,
and S. Jochim,
Phys. Rev. Lett. {\bf{108}}, 075303 (2012).

\bibitem{blume03}
B. Borca, D. Blume, and C. H. Greene,
New J. Phys. {\bf{5}}, 111 (2003).

\bibitem{bohn13}
A. G. Sykes, J. P. Corson, J. P. D'Incao, A. P. Koller, 
C. H. Greene, A. M. Rey, K. R. A. Hazzard, and J. L. Bohn,
Phys. Rev. A {\bf{89}}, 021601(R) (2014).


\bibitem{HuiPRA}
X.-J. Liu, H. Hu, and P. D. Drummond,
Phys. Rev. B {\bf{82}}, 054524 (2010)

\bibitem{Gharashi12}
S.~E.~Gharashi, K.~M.~Daily, and D.~Blume,
Phys. Rev. A {\bf{86}}, 042702 (2012).

\bibitem{Gharashi13}
S.~E.~Gharashi and D.~Blume,
Phys. Rev. Lett. {\bf{111}}, 045302 (2013).


\bibitem{rmp11}
J. Dalibard, F. Gerbier, G. Juzeli\=unas, and P. \"Ohberg,
Rev. Mod. Phys. {\bf{83}}, 1523 (2011).

\bibitem{spielmannat13}
V. Galitski	 and I. B. Spielman,
Nature {\bf{494}}, 49 (2013).

\bibitem{spielmanreview13}
N. Goldman, G. Juzeli\=unas, P. \"Ohberg, and I. B. Spielman,
Preprint at arXiv:1308.6533.

\bibitem{zhaireview}
H. Zhai, 
Int. J. Mod. Phys. B {\bf{26}}, 1230001 (2012).

\bibitem{spielman11}
Y.-J. Lin, K. Jim\'enez-Garc\'ia, and I. B. Spielman,
Nature {\bf{471}}, 83 (2011).

\bibitem{spielman13}
R. A. Williams, M. C. Beeler, L. J. LeBlanc, K. Jim\'enez-Garc\'ia,
and I. B. Spielman, Phys. Rev. Lett. {\bf{111}}, 095301 (2013).

\bibitem{jzhang12}
P. Wang, Z.-Q. Yu, Z. Fu, J. Miao, L. Huang, S. Chai, H. Zhai,
and J. Zhang, Phys. Rev. Lett. {\bf{109}}, 095301 (2012).

\bibitem{jzhang13}
Z. Fu, L. Huang, Z. Meng, P. Wang, L. Zhang, S. Zhang, H. Zhai,
P. Zhang, and J. Zhang,
Nat. Phys. {\bf{10}}, 110 (2014).

\bibitem{zwierlein12}
L. W. Cheuk, A. T. Sommer, Z. Hadzibabic,
T. Yefsah, W. S. Bakr, and M. W. Zwierlein,
Phys. Rev. Lett. {\bf{109}}, 095302 (2012).

\bibitem{engels13}
C. Qu, C. Hamner, M. Gong, C. Zhang, and P. Engels,
Phys. Rev. A {\bf{88}}, 021604(R) (2013).

\bibitem{greene13}
A. Olson, S.-J. Wang, R. J. Niffenegger, C.-H. Li, C. H. Greene, and Y. P. Chen,
Preprint at arXiv:1310.1818.

\bibitem{magarill06}
A. V. Chaplik and L. I. Magarill,
Phys. Rev. Lett. {\bf{96}}, 126402 (2006).

\bibitem{shenoy11}
J. P. Vyasanakere and V. B. Shenoy,
Phys. Rev. B {\bf{83}}, 094515 (2011).

\bibitem{galitski12}
S. Takei, C.-H. Lin, B. M. Anderson, and V. Galitski,
Phys. Rev. A {\bf{85}}, 023626 (2012).

\bibitem{sthbypu}
L. Dong, L. Jiang, H. Hu, and H. Pu,
Phys. Rev. A {\bf{87}}, 043616 (2013).



\bibitem{zinner12}
O. V. Marchuov, A. G. Volosniev, D. V. Fedorov, A. S. Jensen,
and N. T. Zinner,
J. Phys. B {\bf{46}}, 134012 (2012).

\bibitem{anderson2013}
B. M. Anderson and C. W. Clark,
J. Phys. B {\bf{46}}, 134003 (2013).


\bibitem{desmondSOC}
X. Y. Yin, S. Gopalakrishnan, and D. Blume,
Phys. Rev. A {\bf{89}}, 033606 (2014).

\bibitem{puPRA13}
B. Ramachandhran, H. Hu, and H. Pu,
Phys. Rev. A {\bf{87}}, 033627 (2013).


\bibitem{Rashba}
Y. A. Bychkov and E. I. Rashba,
J. Phys. C {\bf{17}}, 6039 (1984).

\bibitem{Dresselhaus}
G. Dresselhaus,
Phys. Rev. {\bf{100}}, 580 (1955).


\bibitem{Olsh98}
M.~Olshanii,
Phys. Rev. Lett. {\bf{81}}, 938 (1998).


\bibitem{olshanii2}
T. Bergeman, M. G. Moore, M. Olshanii,
Phys. Rev. Lett. {\bf{91}}, 163201 (2003).


\bibitem{chinesepaper}
R. Zhang and W. Zhang,
Phys. Rev.A {\bf{88}}, 053605 (2013).

\bibitem{sakurai}
See, e.g., J. J. Sakurai, {\em{Modern Quantum Mechanics}}, Revised Edition,
Addison Wesley, or other quantum texts.


\bibitem{integraltable}
See entry 7.374.7 of I. S. Gradshteyn and I. M. Ryzhik,
{\em{Table of integrals, series, and products}},
6th Ed., Academic Press.



\end{thebibliography}
\end{document}